# Third Order Perturbed Heisenberg Hamiltonian of Thick Spinel Ferrite Films

P. Samarasekara

Department of Physics, University of Peradeniya, Peradeniya, Sri Lanka.

## Abstract

The third order perturbed Heisenberg Hamiltonian was employed to investigate the spinel thick nickel ferrite films. The variation of energy up to N=10000 was studied. At N=75, the energy required to rotate from easy to hard direction is very small. For film with N=10000, the first major maximum and minimum can be observed at $202^0$ and $317^0$, respectively. This curve shows some abrupt changes after introducing third order perturbation. The maximum energy of this curve is higher than that of spinel thick ferrite films with second order perturbed Heisenberg Hamiltonian. At some values of stress induced anisotropy, the maximum energy is less than that of spinel thick ferrite films with second order perturbed Heisenberg Hamiltonian derived by us previously.

## 1. Introduction:

The effect of third order perturbation on the Heisenberg Hamiltonian of spinel ferrite thick films will be described in detail in this report. Previously, the structure of spinel ferrites with the position of octahedral and tetrahedral sites is given in detail [1-5]. Only the occupied octahedral and tetrahedral sited were used for the calculation in this report although there are many filled and vacant octahedral and tetrahedral sites in cubic spinel cell [1]. Few previous reports could be found on the theoretical works of ferrites [6-8]. The solution of Heisenberg ferrites only with spin exchange interaction term has been found earlier by means of the retarded Green function equations [6].

All the relevant energy terms such as spin exchange energy, dipole energy, second and fourth order anisotropy terms, interaction with magnetic field and stress induced anisotropy in Heisenberg Hamiltonian were taken into consideration. These equations derived here can be applied for spinel ferrites with unit cell $AFe_2O_4$ such as $Fe_3O_4$, $NiFe_2O_4$ and $ZnFe_2O_4$ only. The spin exchange interaction energy and dipole interaction have been calculated only between two nearest spin layers and within same spin plane. Also the angle within one cubic cell is assumed to be constant. The change of angle at the interface of cubic cell will be considered. Earlier the second order perturbed Heisenberg Hamiltonian of spinel ferrite thick films[9] and third order perturbed Heisenberg

Hamiltonian of spinel ferrite thin films[10] have been studied. Heisenberg Hamiltonian was employed to find the energy of spinel ferrite films [12, 17, 18] and energy of ferromagnetic films [13, 17, 11]. According to our previous experimental studies, the stress induce anisotropy plays a major role in sputtered ferromagnetic and ferrite thin films [14, 15, 16, 19].

**2. Model:**

The Heisenberg Hamiltonian of a thin film can be written as following.

$$H = -J\sum_{m,n}\vec{S}_m.\vec{S}_n + \omega\sum_{m\neq n}(\frac{\vec{S}_m.\vec{S}_n}{r_{mn}^3} - \frac{3(\vec{S}_m.\vec{r}_{mn})(\vec{r}_{mn}.\vec{S}_n)}{r_{mn}^5}) - \sum_m D_{\lambda_m}^{(2)}(S_m^z)^2 - \sum_m D_{\lambda_m}^{(4)}(S_m^z)^4$$

$$-\sum_m \vec{H}..\vec{S}_m - \sum_m K_s Sin2\theta_m \qquad (1)$$

Here J, ω, θ, $D_m^{(2)}, D_m^{(4)}, H_{in}, H_{out}, K_s$, m, n and N are spin exchange interaction, strength of long range dipole interaction, azimuthal angle of spin, second and fourth order anisotropy constants, in plane and out of plane applied magnetic fields, stress induced anisotropy constant, spin plane indices and total number of layers in film, respectively. When the stress applies normal to the film plane, the angle between m$^{th}$ spin and the stress is $\theta_m$.

The cubic cell has been divided into 8 spin layers with alternative A and Fe spins layers. The spins of A and Fe will be taken as 1 and p, respectively. While the spins in one layer point in one direction, spins in adjacent layers point in opposite directions. A thin film with (001) spinel cubic cell orientation will be considered. The length of one side of unit cell will be taken as "a". Within the cell the spins orient in one direction due to the super exchange interaction between spins (or magnetic moments). Therefore the results proven for oriented case in one of our early report will be used for following equations[13]. But the angle θ will vary from $\theta_m$ to $\theta_{m+1}$ at the interface between two cells. For a thin film with thickness Na,

Spin exchange interaction energy=$E_{exchange}$= N(-10J+72Jp-22Jp$^2$)+8Jp$\sum_{m=1}^{N-1}\cos(\theta_{m+1}-\theta_m)$

Dipole interaction energy=$E_{dipole}$

$$E_{dipole} = -48.415\omega\sum_{m=1}^{N}(1+3\cos 2\theta_m) + 20.41\omega p\sum_{m=1}^{N-1}[\cos(\theta_{m+1}-\theta_m) + 3\cos(\theta_{m+1}+\theta_m)]$$

Here the first and second term in each above equation represent the variation of energy within the cell and the interface of the cell, respectively. Then total energy is given by

$$E = N(-10J+72Jp-22Jp^2) + 8Jp\sum_{m=1}^{N-1}\cos(\theta_{m+1}-\theta_m)$$

$$-48.415\omega\sum_{m=1}^{N}(1+3\cos 2\theta_m) + 20.41\omega p\sum_{m=1}^{N-1}[\cos(\theta_{m+1}-\theta_m) + 3\cos(\theta_{m+1}+\theta_m)]$$

$$-\sum_{m=1}^{N}[D_m^{(2)}\cos^2\theta_m + D_m^{(4)}\cos^4\theta_m]$$

$$-4(1-p)\sum_{m=1}^{N}[H_{in}\sin\theta_m + H_{out}\cos\theta_m + K_s\sin 2\theta_m] \qquad (2)$$

Here the anisotropy energy term and the last term have been explained in our previous report for oriented spinel ferrite. If the angle is given by $\theta_m=\theta+\varepsilon_m$ with perturbation $\varepsilon_m$, after taking the terms up to third order perturbation of $\varepsilon$,

The total energy can be given as $E(\theta)=E_0+E(\varepsilon)+E(\varepsilon^2)+E(\varepsilon^3)$

Here

$E_0 = -10JN+72pNJ-22Jp^2N+8Jp(N-1)-48.415\omega N-145.245\omega N\cos(2\theta)$

$\qquad +20.41\omega p[(N-1)+3(N-1)\cos(2\theta)]$

$$-\cos^2\theta\sum_{m=1}^{N}D_m^{(2)} - \cos^4\theta\sum_{m=1}^{N}D_m^{(4)} - 4(1-p)N(H_{in}\sin\theta + H_{out}\cos\theta + K_s\sin 2\theta) \quad (3)$$

$$E(\varepsilon) = 290.5\omega\sin(2\theta)\sum_{m=1}^{N}\varepsilon_m - 61.23\omega p\sin(2\theta)\sum_{m=1}^{N-1}(\varepsilon_m+\varepsilon_n)$$

$$+\sin 2\theta\sum_{m=1}^{N}D_m^{(2)}\varepsilon_m + 2\cos^2\theta\sin 2\theta\sum_{m=1}^{N}D_m^{(4)}\varepsilon_m$$

$$+4(1-p)[-H_{in}\cos\theta\sum_{m=1}^{N}\varepsilon_m + H_{out}\sin\theta\sum_{m=1}^{N}\varepsilon_m - 2K_s\cos 2\theta\sum_{m=1}^{N}\varepsilon_m] \qquad (4)$$

$$E(\varepsilon^2) = -4Jp\sum_{m=1}^{N-1}(\varepsilon_n - \varepsilon_m)^2 + 290.5\omega\cos(2\theta)\sum_{m=1}^{N}\varepsilon_m^2 - 10.2\omega p\sum_{m=1}^{N-1}(\varepsilon_n - \varepsilon_m)^2$$

$$- 30.6\omega p\cos(2\theta)\sum_{m=1}^{N-1}(\varepsilon_n + \varepsilon_m)^2$$

$$- (\sin^2\theta - \cos^2\theta)\sum_{m=1}^{N}D_m^{(2)}\varepsilon_m^2 + 2\cos^2\theta(\cos^2\theta - 3\sin^2\theta)\sum_{m=1}^{N}D_m^{(4)}\varepsilon_m^2$$

$$+ 4(1-p)[\frac{H_{in}}{2}\sin\theta\sum_{m=1}^{N}\varepsilon_m^2 + \frac{H_{out}}{2}\cos\theta\sum_{m=1}^{N}\varepsilon_m^2 + 2K_s\sin 2\theta\sum_{m=1}^{N}\varepsilon_m^2] \quad (5)$$

$$E(\varepsilon^3) = 10.2 p\omega\sin 2\theta \sum_{m,n=1}^{N}(\varepsilon_m + \varepsilon_n)^3 - 193.66\omega\sin 2\theta\sum_{m=1}^{N}\varepsilon_m^3 - \frac{4}{3}\cos\theta\sin\theta\sum_{m=1}^{N}D_m^{(2)}\varepsilon_m^3$$

$$- 4\cos\theta\sin\theta(\frac{5}{3}\cos^2\theta - \sin^2\theta)\sum_{m=1}^{N}D_m^{(4)}\varepsilon_m^3$$

$$+ 4(1-p)[\frac{H_{in}}{6}\cos\theta\sum_{m=1}^{N}\varepsilon_m^3 - \frac{H_{out}}{6}\sin\theta\sum_{m=1}^{N}\varepsilon_m^3 + \frac{4K_s}{3}\cos 2\theta\sum_{m=1}^{N}\varepsilon_m^3]$$

The sin and cosine terms in equation number 2 have been expanded to obtain above equations. Here n=m+1.

Under the constraint $\sum_{m=1}^{N}\varepsilon_m = 0$, first and last three terms of equation 4 are zero.

Therefore, $E(\varepsilon) = \vec{\alpha}.\vec{\varepsilon}$

Here $\vec{\alpha}(\varepsilon) = \vec{B}(\theta)\sin 2\theta$ are the terms of matrices with

$$B_\lambda(\theta) = -122.46\omega p + D_\lambda^{(2)} + 2D_\lambda^{(4)}\cos^2\theta \quad (6)$$

Also $E(\varepsilon^2) = \frac{1}{2}\vec{\varepsilon}.C.\vec{\varepsilon}$, and matrix C is assumed to be symmetric ($C_{mn}=C_{nm}$).

Here the elements of matrix C can be given as following,

$C_{m, m+1}$=8Jp+20.4ωp-61.2pωcos(2θ)

For m=1 and N,

$C_{mm}$= -8Jp-20.4ωp-61.2pωcos(2θ)+581ωcos(2θ) $- 2(\sin^2\theta - \cos^2\theta) D_m^{(2)}$

$\quad + 4\cos^2\theta(\cos^2\theta - 3\sin^2\theta) D_m^{(4)} + 4(1-p)[H_{in}\sin\theta + H_{out}\cos\theta + 4K_s\sin(2\theta)]$ **(7)**

For m=2, 3, ----, N-1

$C_{mm}$= -16Jp-40.8ωp-122.4pωcos(2θ)+581ωcos(2θ) $- 2(\sin^2\theta - \cos^2\theta) D_m^{(2)}$

$+ 4\cos^2\theta(\cos^2\theta - 3\sin^2\theta) D_m^{(4)} + 4(1-p)[H_{in}\sin\theta + H_{out}\cos\theta + 4K_s\sin(2\theta)]$

Otherwise, $C_{mn}=0$

Also $E(\varepsilon^3) = \varepsilon^2 \beta.\vec{\varepsilon}$

Here matrix elements of matrix β can be given as following.

When m=1 and N,

$$\beta_{mm} = -193.66\omega\sin 2\theta + 10.2 p\omega\sin 2\theta - \frac{4}{3}\cos\theta\sin\theta D_m^{(2)}$$

$$- 4\cos\theta\sin\theta(\frac{5}{3}\cos^2\theta - \sin^2\theta)D_m^{(4)} + 4(1-p)[\frac{H_{in}}{6}\cos\theta - \frac{H_{out}}{6}\sin\theta + \frac{4K_s}{3}\cos 2\theta]$$

When m=2, 3, ------, N-1

$$\beta_{mm} = -193.66\omega\sin 2\theta + 20.4 p\omega\sin 2\theta - \frac{4}{3}\cos\theta\sin\theta D_m^{(2)}$$

$$- 4\cos\theta\sin\theta(\frac{5}{3}\cos^2\theta - \sin^2\theta)D_m^{(4)} + 4(1-p)[\frac{H_{in}}{6}\cos\theta - \frac{H_{out}}{6}\sin\theta + \frac{4K_s}{3}\cos 2\theta]$$

$\beta_{m,m+1} = 30.6 p\omega\sin 2\theta$ **(8)**

Otherwise $\beta_{nm}=0$. Also $\beta_{nm}=\beta_{mn}$ and matrix β is symmetric.

Therefore, the total magnetic energy given in equation 2 can be deduced to

$E(\theta) = E_0 + \vec{\alpha}.\vec{\varepsilon} + \frac{1}{2}\vec{\varepsilon}.C.\vec{\varepsilon} + \varepsilon^2 \beta.\vec{\varepsilon}$ **(9)**

Because the derivation of a final equation for ε with the third order of ε in above equation is tedious, only the second order of ε will be considered for following derivation.

Then $E(\theta) = E_0 + \vec{\alpha}.\vec{\varepsilon} + \frac{1}{2}\vec{\varepsilon}.C.\vec{\varepsilon}$

Using a suitable constraint in above equation, it is possible to show that $\vec{\varepsilon} = -C^+.\vec{\alpha}$

Here $C^+$ is the pseudo-inverse given by

$$C.C^+ = 1 - \frac{E}{N}. \qquad (10)$$

E is the matrix with all elements given by $E_{mn}=1$.

After using ε in equation 9, $E(\theta)=E_0 - \frac{1}{2}\vec{\alpha}.C^+.\vec{\alpha} - (C^+\alpha)^2 \vec{\beta}(C^+\alpha)$ (11)

## 3. Results and discussion:

When N is very large (Ex: N=10000), $CC^+=1$, and $C^+$ is the standard inverse matrix of C. When the difference between m and n is one, $C_{m,\,m+1}$=8Jp+20.4ωp-61.2pωcos(2θ). If $H_{in}$, $H_{out}$ and $K_s$ are very large, then $C_{11} \gg C_{12}$. If this $C_{m,\,m+1}=0$, then the matrix C becomes diagonal, and the elements of inverse matrix $C^+$ is given by $C^+_{mm} = \frac{1}{C_{mm}}$. Therefore all the derivation will be done under above assumption for the convenience.

Then $\alpha_1 = ---- = \alpha_n = [-122.46\omega p + D_\lambda^{(2)} + 2D_\lambda^{(4)} \cos^2\theta]\sin(2\theta)$

$\alpha.C^+.\alpha = 2C^+_{11}\alpha_1^2 + +\alpha_1^2(N-2)C^+_{22} = \frac{2\alpha_1^2}{C_{11}} + \frac{\alpha_1^2(N-2)}{C_{22}}$

For Nickel ferrite with p=2.5,

$E_0$= 52.5JN-20J-48.415ωN-145.245ωNcos(2θ)+51.025ω(N-1)[1+3cos(2θ)]

$\quad - N[\cos^2\theta D_m^{(2)} + \cos^4\theta D_m^{(4)} - 6(H_{in}\sin\theta + H_{out}\cos\theta + K_s\sin 2\theta)]$

$C_{11}=C_{NN}$= -20J-51ω+428ωcos(2θ)+$2\cos 2\theta\, D_m^{(2)} + 4\cos^2\theta(\cos^2\theta - 3\sin^2\theta)\, D_m^{(4)}$

$\quad - 6[H_{in}\sin\theta + H_{out}\cos\theta + 4K_s\sin(2\theta)]$

$C_{22}=C_{33}$=------=$C_{N-1,N-1}$= -40J-102ω+275ωcos(2θ)+$2(\cos 2\theta)\, D_m^{(2)}$

$\quad + 4\cos^2\theta(\cos^2\theta - 3\sin^2\theta)\, D_m^{(4)} - 6[H_{in}\sin\theta + H_{out}\cos\theta + 4K_s\sin(2\theta)]$

$\alpha_1 = [-306.15\omega + D_\lambda^{(2)} + 2D_\lambda^{(4)}\cos^2\theta]\sin(2\theta)$

$\beta_{11} = \beta_{NN} = -168.16\omega\sin 2\theta - \frac{4}{3}\cos\theta\sin\theta D_m^{(2)}$

$\quad - 4\cos\theta\sin\theta(\frac{5}{3}\cos^2\theta - \sin^2\theta)D_m^{(4)} - 6[\frac{H_{in}}{6}\cos\theta - \frac{H_{out}}{6}\sin\theta + \frac{4K_s}{3}\cos 2\theta]$

$$\beta_{22} = -142.66\omega\sin 2\theta - \frac{4}{3}\cos\theta\sin\theta D_m^{(2)}$$

$$- 4\cos\theta\sin\theta(\frac{5}{3}\cos^2\theta - \sin^2\theta)D_m^{(4)} - 6[\frac{H_{in}}{6}\cos\theta - \frac{H_{out}}{6}\sin\theta + \frac{4K_s}{3}\cos 2\theta]$$

$$\beta_{m,m+1} = 76.5\omega\sin 2\theta$$

$$(C^+\alpha)^2\beta(C^+\alpha) = (C_{11}{}^+\alpha_1)^2(\beta_{11}C_{11}{}^+\alpha_1 + \beta_{12}C_{22}{}^+\alpha_2 + \text{-------} + \beta_{1N}C_{NN}{}^+\alpha_N)$$
$$+(C_{22}{}^+\alpha_2)^2(\beta_{21}C_{11}{}^+\alpha_1 + \beta_{22}C_{22}{}^+\alpha_2 + \text{-------} + \beta_{2N}C_{NN}{}^+\alpha_N)$$
$$+(C_{33}{}^+\alpha_3)^2(\beta_{31}C_{11}{}^+\alpha_1 + \beta_{32}C_{22}{}^+\alpha_2 + \text{-------} + \beta_{3N}C_{NN}{}^+\alpha_N) + \text{--------}$$
$$\text{------} + (C_{NN}{}^+\alpha_N)^2(\beta_{N1}C_{11}{}^+\alpha_1 + \beta_{N2}C_{22}{}^+\alpha_2 + \text{-------} + \beta_{NN}C_{NN}{}^+\alpha_N)$$

$$(C^+\alpha)^2\beta(C^+\alpha) = \alpha^3[\frac{2}{C_{11}{}^2}(\frac{\beta_{11}}{C_{11}} + \frac{\beta_{12}}{C_{22}}) + \frac{2}{C_{22}{}^2}(\frac{\beta_{12}}{C_{11}} + \frac{\beta_{22} + \beta_{12}}{C_{22}}) + \frac{N-4}{C_{22}{}^3}(2\beta_{12} + \beta_{22})]$$

The total energy can be found from equation 11. When $\frac{J}{\omega} = \frac{D_m^{(2)}}{\omega} = \frac{H_{in}}{\omega} = \frac{H_{out}}{\omega} = \frac{K_s}{\omega} = 10$, $and\ \frac{D_m^{(4)}}{\omega} = 5$, the 3-D plot of $\frac{E(\theta)}{\omega}$ versus θ and N is given in figure 1. Although the equation is valid for large values of N only, the graph has been drawn for even small values of N too in order to study the variation of energy at small values of N too. The maximum energy of this thick film is almost same as that of spinel thick ferrite films with second order perturbed Heisenberg Hamiltonian[10]. Energy variation of these two is also similar. Near N=75, the separation between maximum and minimum energies are very small implying that the energy required to rotate from easy to hard direction is small at this N value. The maximum energy of this film is higher than that of ferromagnetic thick films with third order perturbation[12]. Energy variation is different from that of ferromagnetic thick films with third order perturbation.

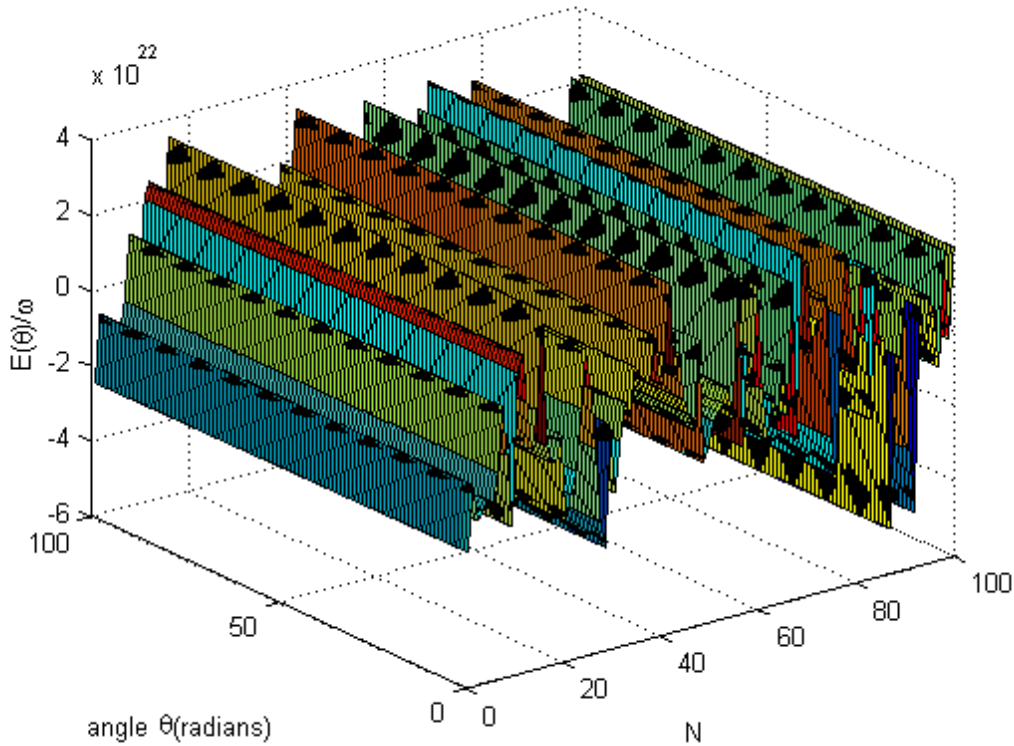

Figure 1: 3-D plot of $\frac{E(\theta)}{\omega}$ versus θ and N for Nickel ferrite

When N=10000, the graph between $\frac{E(\theta)}{\omega}$ and θ is given in figure 2. The maximum energy of this curve is higher than that of spinel thick ferrite films with second order perturbed Heisenberg Hamiltonian[10]. Due to sudden changes, this energy curve is less smoother compared with that of spinel thick ferrite films with second order perturbed Heisenberg Hamiltonian. The first major maximum and minimum can be observed at $202^0$ and $317^0$, respectively. Angle between easy and hard directions is not $90^0$ in this case. Some sudden changes could be observed in ferromagnetic thick ferrite films with third order perturbation[12]. The maximum energy is almost same as that of ferromagnetic thick ferrite films with third order perturbation. Positions of easy and hard directions and separation between easy and hard directions are different from those of ferromagnetic thick ferrite films with third order perturbation and spinel thick ferrite films with second order perturbation.

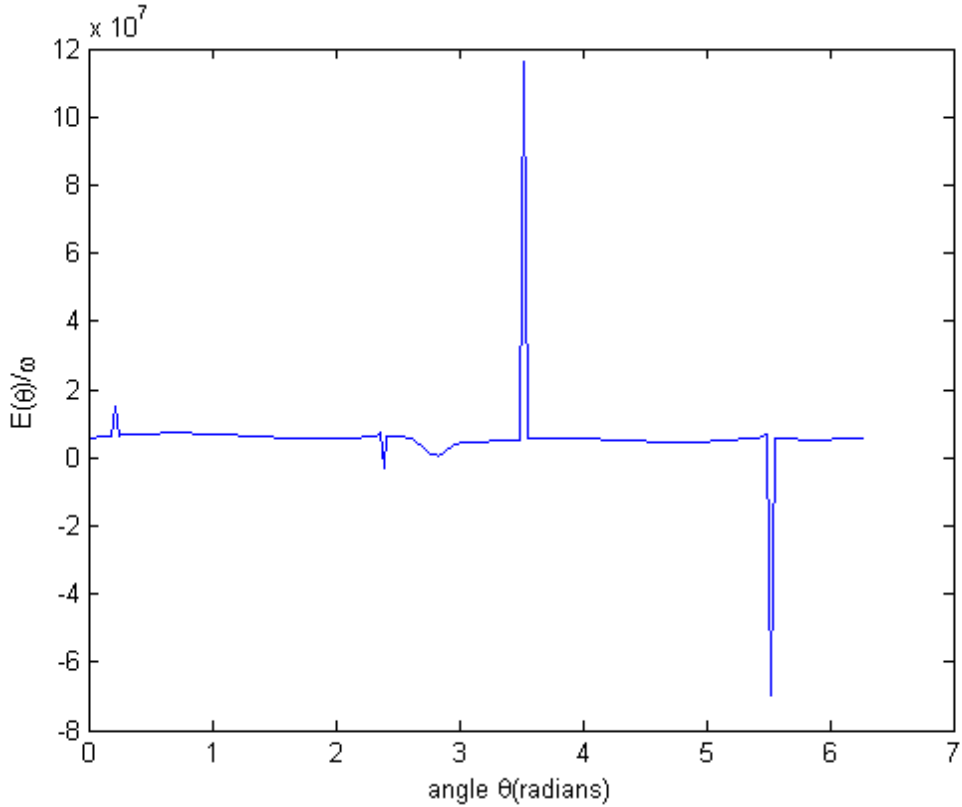

Figure 2: Graph between $\frac{E(\theta)}{\omega}$ and θ for N=10000

When N=10000 and $\frac{K_s}{\omega}$ is a variable, the 3-D plot of $\frac{E(\theta)}{\omega}$ versus θ and $\frac{K_s}{\omega}$ is given in figure 3. The maximum energy of this curve is less than that of spinel thick ferrite films with second order perturbed Heisenberg Hamiltonian[10]. The variation of energy is also different from that of spinel thick ferrite films with second order perturbed Heisenberg Hamiltonian.

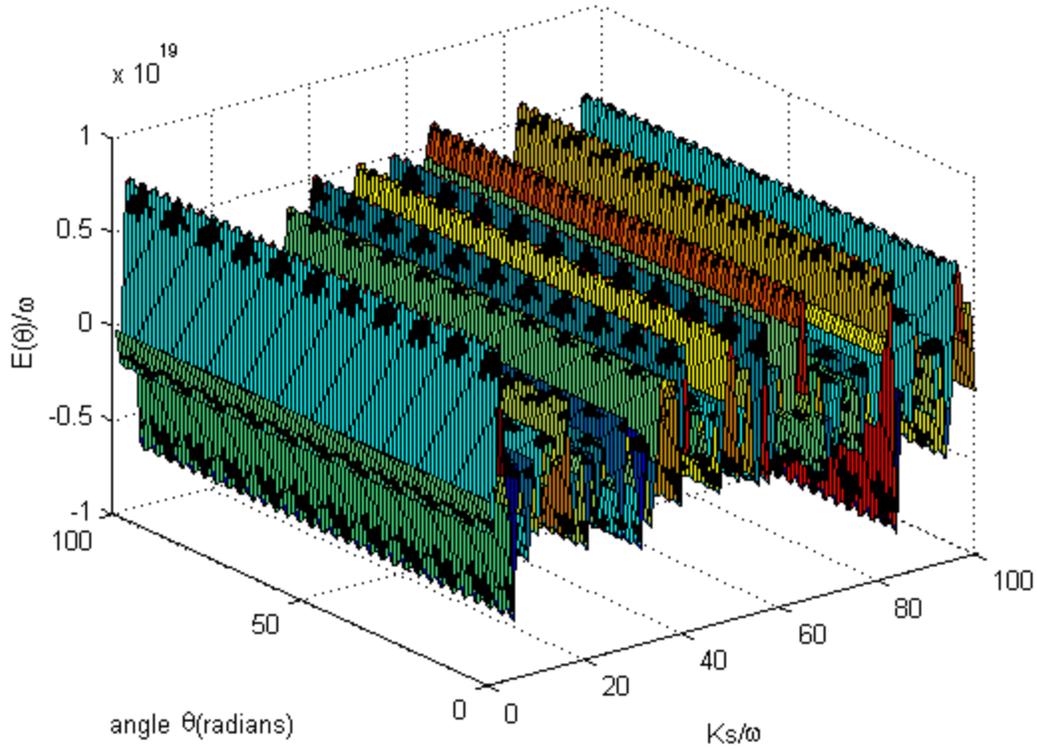

Figure 3: 3-D plot of $\frac{E(\theta)}{\omega}$ versus $\theta$ and $\frac{K_s}{\omega}$ for N=10000

## 4. Conclusion:

The variation of energy with angle, thickness and stress was studied up to N=10000. Near N=75, the energy required to rotate from easy to hard direction is very small implying that the anisotropy energy is small at this value N. For film with N=10000, the major maximum and minimum can be observed at $202^0$ and $317^0$, respectively. Introducing third order perturbation destroys the smoothness of energy curve. The maximum energy of this curve is higher than that of spinel thick ferrite films with second order perturbed Heisenberg Hamiltonian. In 3-D plot of $\frac{E(\theta)}{\omega}$ versus $\theta$ and $\frac{K_s}{\omega}$, some energy minimums can be observed indicating that the film can be oriented in some particular directions by applying certain stresses.